\documentclass[aps,pra,groupedaddress,showpacs,twocolumn]{revtex4}

\usepackage{graphics}
\usepackage{epsfig}
\usepackage{amsmath}

\begin{document}

\title{Exact solution of the ion-laser interaction in all regimes}
\author{ A. Z\'u\~niga-Segundo, }
\affiliation{Departamento de F\'{\i}sica, Escuela Superior de
F\'{\i}sica y Matem\'aticas Edificio 9, Unidad Profesional 'Adolfo
L´opez Mateos', 07738 M\'exico, DF, Mexico}
\author{J.M. Vargas-Mart\'{\i}nez}
\affiliation{Benem\'erita Univ. Aut\'onoma Puebla, Fac Ciencias
F\'{\i}s. Mat., Apartado Postal 1152, Puebla, Pue. 72000, Mexico}
\author{ R. Ju\'arez-Amaro,}
\affiliation{Universidad Tecnol\'ogica de la Mixteca, Apdo. Postal
71, 69000 Huajuapan de Le\'on, Oax., Mexico}
\author{  H. Moya-Cessa}
\affiliation{ INAOE, Coordinaci\'on de Optica, Apdo. Postal 51 y
216, 72000 Puebla, Pue., Mexico}

\begin{abstract}
 We show that in the trapped ion-laser interaction all the
 regimes may be considered analytically. We may solve not only for
 different laser intensities, but also away from resonance and
 from the Lamb-Dicke regime. It is found a dispersive Hamiltonian
 for the high intensity regime, that, being diagonal, its evolution operator
 may be easily calculated.
\end{abstract}
 \maketitle

\section{Introduction}

The ion-laser interaction may be easily solved in the low
intensity regime (LIR)
\cite{wine,wine2,wine3,ion,Wall,Buzek,tomb}, but besides the
condition that the laser intensity is much lower than the
vibrational frequency, we set the condition that the detuning
between the laser and the atomic transition frequency is an
integer multiple  of the vibrational frequency. Then some
questions arise: Is it possible not to consider integer multiples
of the vibrational frequency? Is it possible to solve for high and
middle intensities?

Indeed, Moya-Cessa {\it et al.} \cite{moya} have shown that it is
possible to find solutions for any set of parameters, i.e. in all
the regimes. However the solutions are not general because the set
of eigenstates found can not expand all possible (general) states.

It has been shown already that for low intensities it is possible
also to consider the ion micromotion \cite{Schleich}, and by using
Ermakov-Lewis invariant methods \cite{Manuel} it was possible to
{\it linearize} the ion-laser Hamiltonian when the micromotion was
included \cite{Jose}. Here we would like to show how it is
possible to solve the interaction in different regimes, including
high intensity and medium intensity. the method allows also not to
consider multiple integers of the vibrational frequency.
\section{Ion-laser interaction}

The Hamiltonian for the ion-laser dipole interaction, with no
approximations can be written as
\begin{eqnarray} \nonumber
\hat{H}&=&\nu \hat{n}+\frac{\omega_{a} }{2}\hat{\sigma} _{z}+
\Omega \left( \hat{\sigma}_{+} +\hat{\sigma}_{-}\right)\\ &\times
& \left(e^{[i\alpha (a+a^{\dagger})-\omega_{L} t]}+e^{-i[\alpha
(a+a^{\dagger})-\omega_{L} t]}\right), \label{ion}
\end{eqnarray}
where $\nu $ is the harmonic trapping frequency, $\omega_{a}$ is
the atomic transition frequency, $\omega_{L}$ is the field
frequency, $\Omega $ the (real) Rabi frequency of the ion-laser
coupling and $\eta $ the Lamb-Dicke parameter. The operators
$a^{\dagger}$ and $a$ are the creation and annihilation operators
for the vibrational motion of the ion, and the $\sigma$'s are the
Pauli spin operators.

By doing the transformation $\hat{T}\vert\psi\rangle$ with
$\hat{T}=\exp(-i\frac{\omega_{a}+\delta }{2}\hat{\sigma} _{z}t)$
and performing the optical RWA \cite{foot} we arrive at the
well-known Hamiltonian
\begin{figure}[hbt]
\begin{center}
\includegraphics[width=0.35\textwidth]{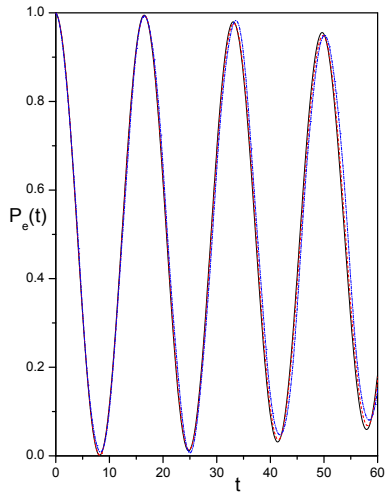}
\end{center}
\caption{\label{fig1} Plot of $P_e(t)$ as a function of $t$ for
$k=0$, $\nu=1$, $\Omega=0.2$ and $\eta=0.1$. Solid line represents
the numerical (exact) solution, dashed line the solution from
Section II and the dot-dashed line the solution for the dispersive
Hamiltonian of Section III.}
\end{figure}
\begin{equation}
\hat{H}_{ion}=\nu \hat{n}+\frac{\delta }{2}\hat{\sigma} _{z}+
\Omega \left( \hat{\sigma}_{+} \hat{D}(i\eta)+\hat{\sigma}_{-}
\hat{D}^{\dagger}(i\eta)\right), \label{ion1}
\end{equation}
where $\hat{D}(i\alpha) = e^{i\alpha (a+a^{\dagger})}$ is the
displacement operator, and
 $\delta =\omega _{a}-\omega _{L}$ the laser-ion detuning.
\subsection{Low intensity regime} The low intensity regime is the
well-known regime, where several effects like multi-phonon
transitions, Jaynes-Cummings (JC)
 and anti-JC interactions may be obtained. To solve this regime,
 we follow first the traditional approach. We start by using
 Baker-Hausdorff formula \cite{Louisel} to factor the displacement
 operators in equation (2) into a product of exponentials
 and consider $\delta=k\nu$, i.e. an integer multiple of
 $\nu$ with $k=0,\pm 1, \pm 2, ...$, we then obtain
\begin{equation}
\hat{H}_{ion}=\nu \hat{n}+\frac{k\nu }{2}\hat{\sigma} _{z}+ \Omega
e^{-\eta^2/2} \left( \hat{\sigma}_{+} e^{i\eta
a^{\dagger}}e^{i\eta a}+\hat{\sigma}_{-} e^{-i\eta
a^{\dagger}}e^{-i\eta a}\right).
\end{equation}
Now we expand the exponentials of the annihilation and creation
operators in Taylor series and get rid off the free Hamiltonians
via a transformation to the interaction picture to obtain the
Hamiltonian
\begin{eqnarray}
\hat{H}_{I}&=& \Omega e^{-\eta^2/2}\\ \nonumber & \times & \left(
\hat{\sigma}_{-} \sum_{n,m=0}^{\infty}
\frac{(-i\eta)^{n+m}}{n!m!}a^{\dagger n}a^m e^{i\nu t(n-m+k)}+H.c
\right).
\end{eqnarray}
We use the fact that are in the LIR, $\nu\gg\Omega$ and make the
RWA, i.e. we only keep time independent terms in the above
Hamiltonian to end up with
\begin{equation}
\hat{H}_{I}= \Omega e^{-\eta^2/2} \left(a^{\dagger k}(-i\eta)^{k}
\hat{\sigma}_{-}
\frac{\hat{n}!}{(\hat{n}+k)!}L_{\hat{n}}^{(k)}(\eta^2)+H.c
\right), \label{trad}
\end{equation}
with $L_{\hat{n}}^{(k)}(x)$ the associated Laguerre polynomials of
order  (operator) $\hat{n}=a^{\dagger}a$. The Hamiltonian above is
now readily solvable, so that we may find easily the evolution
operator associated to it.
\begin{figure}[hbt]
\begin{center}
\includegraphics[width=0.35\textwidth]{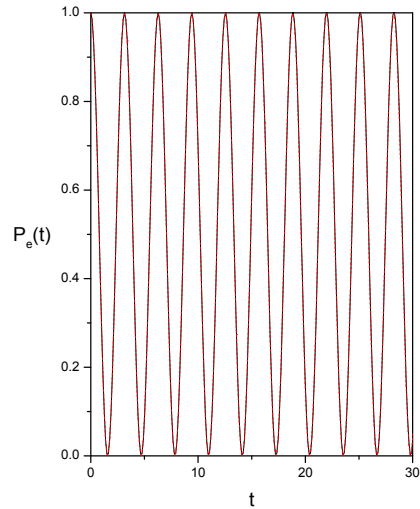}
\end{center}
\caption{\label{fig2} Plot of $P_e(t)$ as a function of $t$ for
$k=0$, $\Omega=1$, $\nu=0.2$ and $\eta=0.1$. Solid line represents
the numerical (exact) solution, dashed line the solution for the
dispersive Hamiltonian of  Section III .}
\end{figure}

\section{Other regimes}
 Although the atom-field and ion-laser
interactions appear to be physically and mathematically quite
distinct, they are in fact exactly equivalent. The easiest way to
see this is by using the transformation
\begin{equation} \hat{R}=e^{i\hat{n}\frac{\pi}{2}}e^{\frac{\pi}{4}(\hat{\sigma}_+-\hat{\sigma}_-)}
e^{-i\frac{\eta}{2}(\hat{a}+\hat{a}^{\dagger})(\hat{\sigma_z})}
\end{equation}
such that $\hat{{\mathcal
H}}_{ion}=\hat{R}\hat{H}_{ion}\hat{R}^{\dagger}$
\begin{equation}
\hat{{\mathcal H}}_{ion} =\nu \hat{n}+\Omega \hat{\sigma} _{z}+\frac{ \eta \nu }{2}%
\left( \hat{\sigma} _{+}+\hat{\sigma} _{-}\right) \left( \hat{a}+\hat{a}^{\dag }\right) +\frac{%
\delta }{2}\left( \hat{\sigma} _{+}+\hat{\sigma} _{-}\right)
+\frac{\nu \eta ^{2}}{4}. \label{JCM2}
\end{equation}

Therefore we have {\it linearized} the ion-laser interaction in an
exact way. In the following we will neglect the term  $\frac{\nu
\eta ^{2}}{4}$ because it only represents a constant shift of all
the eigenenergies.

\subsection{Medium intensity regime (MIR)}
We now consider the case where the vibrational frequency is of the
order of (twice) the field intensity (Rabi frequency). We also
consider the Lamb-Dicke regime, i.e. $\eta \ll  1$. For simplicity
we will set $\delta=0$ to show the different possibilities we have
now. However it is not difficult to produce effective Hamiltonians
also in the off-resonance case. In this case the Hamiltonian
(\ref{JCM2}) may be casted into
\begin{equation}
\hat{{\mathcal H}}_{MIR} =\nu \hat{n}+\Omega \hat{\sigma} _{z}+\frac{ \eta \nu }{2}%
\left( \hat{\sigma} _{+}\hat{a}++\hat{a}^{\dagger }\hat{\sigma}
_{-}\right)
\end{equation}
which is a Hamiltonian that has been extensively studied
\cite{JCM,JSM}.
\subsection{Low and high intensity regimes (HIR)}
We have shown in Section II how  to solve for the LIR case. Here
we will show a different method that is also valid for the HIR.

By transforming the Hamiltonian (\ref{JCM2}) with the unitary
operators
\begin{equation}
\hat{U}_1=
e^{\xi_1(\hat{a}^{\dagger}\hat{\sigma}_+-\hat{a}\hat{\sigma}_-)},
\qquad \hat{U}_2=
e^{\xi_2(\hat{a}\hat{\sigma}_+-\hat{a}^{\dagger}\hat{\sigma}_-)},
\end{equation}
with $\xi_1,\xi_2 \ll 1$, we can remain up to first order in the
expansion $e^{\xi A}Be^{-\xi A}=B+\xi
[A,B]+\frac{\xi^2}{2!}[A,[A,B]]+ ...\approx B+\xi [A,B]$, so we
obtain the effective Hamiltonian 
\begin{eqnarray} \nonumber
\hat{{\mathcal H}}_{eff}&=& \nu \hat{a}^{\dagger} \hat{a} + \Omega
\hat{\sigma}_z - \chi_{ion} \hat{\sigma}_z (\hat{a}^{\dagger}
\hat{a}+\frac{1}{2})+\frac{\delta}{2}(\sigma_++\sigma_-)\\& +&
\frac{\kappa }{2}\hat{\sigma}_{z}\left( \hat{a}^{\dagger
}+\hat{a}\right).\label{effion}
\end{eqnarray}
We have used
\begin{equation}
\xi_1=\frac{\eta\nu}{2(\nu+2\Omega)}\qquad
\xi_2=\frac{\eta\nu}{2(2\Omega-\nu)}.
\end{equation}
We can see that in fact $\xi_1,\xi_2 \ll 1$ either in the LIR (in
this case we have also to consider $\eta \ll 1$) or in the HIR (no
constrain for $\eta$), which justifies completely the
approximation for the above Hamiltonian. For the resonant case,
$k=0$, it becomes diagonal and we can solve it in an easy way. In
Fig.~1 we show a plot for the probability to find the ion in its
excited state. The three curves in the figure correspond to the
exact case (solid line), the solution form Hamiltonian
(\ref{trad}) (dashed line) and the solution for the dispersive
Hamiltonian (\ref{effion}). We can see excellent agreement among
the  three plots for the LIR. Now, for the HIR we show a plot in
Fig.~2 for the numerical solution (solid line) and our solution
from this Section (dashed line). Again it may be noticed an
excellent agreement between both curves. We should stress that
there is no other analytical solution to compare with, as ours is
the first analytical solution in this regime (also in the medium
regime).

 The new interaction constants in the effective
Hamiltonian (\ref{effion}) have the form
\begin{equation}
\chi_{ion} =\frac{2\eta ^{2}\nu ^{2}\Omega }{4\Omega ^{2}-\nu
^{2}},\quad \kappa =\frac{\delta \eta \nu ^{2}}{4\Omega ^{2}-\nu
^{2}}.
\end{equation}

In the resonant case and high intensity regime, $\Omega \gg \nu$,
it is easy to show that
\begin{equation}
\chi_{ion} \rightarrow \chi^{high} =\frac{2\eta ^{2}\nu
^{2}}{4\Omega }\frac{1}{1-\frac{\nu^{2}}{4\Omega^{2} } } \approx
\frac{\eta^2\nu^2}{2\Omega}.
\end{equation}

while in the low intensity regime, $\Omega \ll \nu$, we will have
the same Hamiltonian but $\chi$ will change to
\begin{equation}
\chi_{ion} \rightarrow \chi_{low} = -2\eta^2 \Omega
\frac{1}{1-\frac{4\Omega^2}{\nu^2}}\approx -2\eta^2\Omega
\end{equation}

If in Eq. (\ref{effion}) we take the detuning $\delta$ different
from zero, we could get the usual blue and red side-bands
interactions (see for instance Ref.
 \cite{tomb}). This is done by choosing the value  $\delta=\pm \nu$. The only case in
which we can obtain such regimes is the low intensity case, where
one can perform the RWA to the Hamiltonian (\ref{effion}), which
agrees with the usual procedure for obtaining such blue and red
side-band regimes. The high intensity case, $\Omega \gg \nu $ does
not allow such side-bands because in the Hamiltonian
(\ref{effion}) the interaction constants multiplying the different
terms may be of the same order.
\section{Conclusions}
We have shown that it is possible to solve analytically the
ion-laser Hamiltonian in different intensity regimes, from low to
high. For the MIR we have casted the ion laser Hamiltonian into a
JCM Hamiltonian (for the on-resonant case) that allows easy
solution. For the HIR we have found a dispersive Hamiltonian,
which, being diagonal, it is direct to solve. We have found
excellent agreement between the exact (numerical) solutions and
our proposed solutions.

We would like to thank CONACYT for support.


\begin{thebibliography}{99}
\bibitem{foot} We can perform the
optical rotating wave approximation as the laser and atomic
transition frequencies are much, much larger than the field
intensity in experiments \cite{ion}.
\end{thebibliography}
\end{document}